\documentclass{desyproc}
\newcommand\ignore[1]{}

\newcommand\be{\begin{equation}}
\newcommand\ee{\end{equation}}
\newcommand\bea{\begin{eqnarray}}
\newcommand\eea{\end{eqnarray}}

\newcommand{\bdm}{\begin{displaymath}}
\newcommand{\edm}{\end{displaymath}}
\newcommand\nn{ \nonumber\\}

\renewcommand{\>}{\rangle}

\begin{document}
\title{Holographic Double Diffractive Production of Higgs\\
and the AdS Graviton/Pomeron}


\author{{\slshape Richard Brower$^1$, Marko Djuri\'c$^2$, Chung-I Tan$^3$}\\[1ex]
$^1$Physics Department, Boston University, Boston, MA  02215, USA.\\
$^2$Centro de F\'isica do Porto, Universidade do Porto, 4169-007 Porto, Portugal.\\
$^3$Physics Department, Brown University, Providence, RI 02912, USA.}

\contribID{smith\_joe}


\maketitle

\begin{abstract}
 The holographic approach to double diffractive Higgs production is presented in terms of exchanging the AdS graviton/Pomeron. The goal is to provide a simple  framework  for central exclusive production from a dual strong coupling perspective. 
 \end{abstract}

\section{Introduction}

\label{sec:intro}

A promising production mechanism for Higgs meson at the LHC involves the forward proton-proton scattering $p p \rightarrow p H p$. The protons scatter through very
small angles with a large rapidity gaps separating the Higgs in the
central region. 
Current phenomenological treatment for  the diffractive Higgs production
cross section have generally followed two approaches:  perturbative (weak coupling) vs  confining (strong coupling), or equivalently, in terms of the Regge language,  often referred to as the ``hard Pomeron''  vs``soft Pomeron''  methods. The Regge approach to high energy scattering, although well motivated phenomenologically, has suffered in the past by the lack of a precise theoretical underpinning. The advent of AdS/CFT has dramatically changed the situation. In a holographic approach,  the Pomeron is a well-defined concept and it can be identified as the ``AdS graviton" in the strong coupling~\cite{Brower:2006ea}, or, simply the BPST Pomeron. We briefly review here the general properties of the BPST Pomeron and show how it can be used to describe central double-diffractive particle production, and, in particular,  for  Higgs.

\begin{figure}[h]
\qquad
\includegraphics[height=0.15 \textwidth,width=0.4\textwidth]{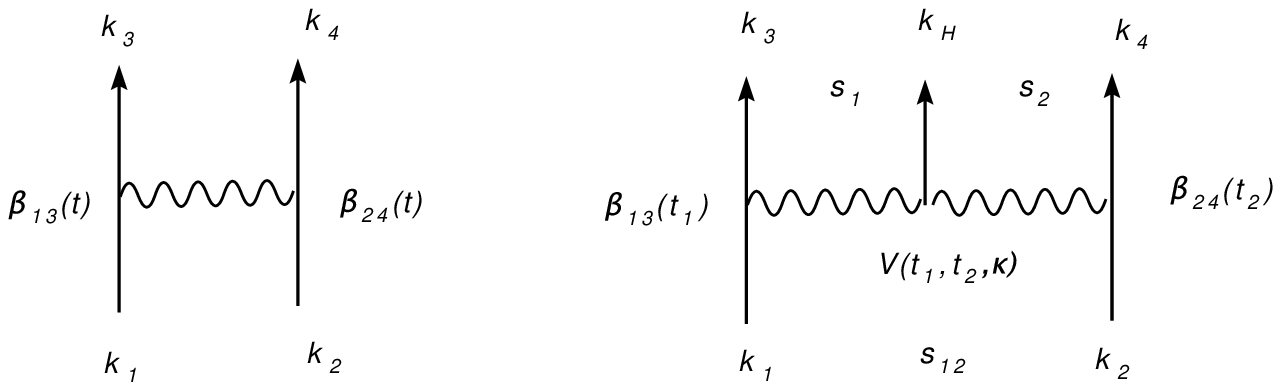}
\qquad
\qquad
\qquad
\includegraphics[angle = 90, height = 0.15\textwidth, width = 0.25\textwidth]{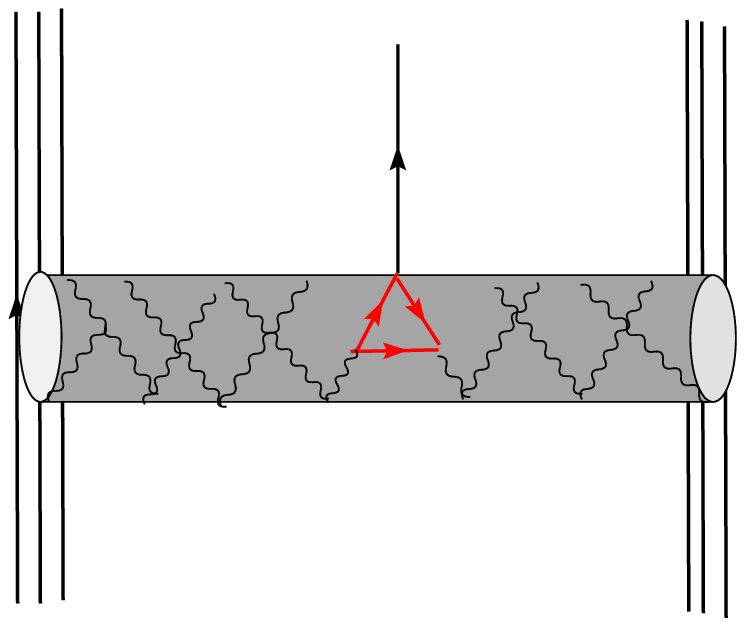}

 \caption{(a) Kinematics for single-Regge limit for  2-to-2 amplitudes, (b) Double-Regge kinematics for  2-to-3 amplitudes.  (c) Cylinder Diagram for large $N_c$ Higgs Production.}
 \label{fig:cylindarHiggs}
\end{figure}

The formulation of AdS/CFT  for high energy diffractive collision
 has already a rather extensive literature to draw on~\cite{Brower:2007qh,Brower:2007xg,Cornalba:2006xm}.
``Factorization in AdS" has emerged as a {\it universal} feature, applicable to scattering involving both particles and currents.  For instance, for elastic scattering, the amplitude can be represented schematically in a factorizable form, 
\be
A(s,t)  = \Phi_{13}*\widetilde {\cal K}_P * \Phi_{24} \; .
\label{eq:adsPomeronScheme}
\ee
where $ \Phi_{13}$ and $ \Phi_{24}$ represented two elastic vertex couplings, and $\widetilde {\cal K}_P$ is an universal Pomeron kernel, with a characteristic power behavior at large $s>>|t|$, 
$
\widetilde {\cal K}_P\sim s^{j_0}\, ,
\label{eq:Regge}
$
schematically represented by Fig. \ref{fig:cylindarHiggs}a. This ``Pomeron intercept", $j_0$, lies in the range  $1<j_0<2$ and is a function of the 't Hooft coupling, $\lambda=g^2N_c$. The  convolution  in (\ref{eq:adsPomeronScheme}), denoted by the $*$-operation, involves an integration over the AdS location in the bulk.     In moving from elastic to the small-$x$ limit for the  deep inelastic scattering (DIS)~\cite{Brower:2010wf}, one simply replaces $\Phi_{13}$ in   (\ref{eq:adsPomeronScheme}) by appropriate product of propagators  for external currents~\cite{Polchinski:2002jw,Brower:2010wf}.

A holographic treatment of central diffractive production  amounts to a generalization of our previous $AdS$ treatment for 2-to-2 amplitudes to one for  2-to-3 amplitudes~\cite{Herzog:2008mu}, e.g., from Fig. \ref{fig:cylindarHiggs}a to Fig. \ref{fig:cylindarHiggs}b.  In this analysis, one generalizes (\ref{eq:adsPomeronScheme}) to 2-to-3 amplitude where
\be
A(s,s_1, s_2, t_1, t_2)  = \Phi_{13}*  \widetilde{\cal K}_P*V_H*  \widetilde {\cal K}_P* \Phi_{24} \; ,
\label{eq:adsDoublePomeronScheme}
\ee
schematically represented by Fig. \ref{fig:cylindarHiggs}b.   A more refined analysis for Higgs production requires a careful treatment for that depicted in Fig. \ref{fig:cylindarHiggs}c. 
A new aspect, not addressed in \cite{Herzog:2008mu},  is the issue of scale invariance breaking.   A proper accounting for a non-vanishing gluon condensate $\langle F^2\rangle$ turns out to be a crucial ingredient in understanding the strength of diffractive Higgs production~\cite{Kharzeev:2000jwa,Brower:2012mk}.

 The basic theoretical steps necessary in order to arrive at (\ref{eq:adsPomeronScheme}) and (\ref{eq:adsDoublePomeronScheme}) are: 

(a)  {\it \bf Diffractive Scattering and  Large $N_c$ Limit}: In leading order of the $1/N_c$
expansion at fixed 't Hooft coupling $\lambda = g^2N_c$, diffraction is
given peturbatively by the exchange of a network of gluons with the
topology of a cylinder, corresponding in a confining theory to the
t-channel exchange of a closed string for glueball states.  Such a state can be  identified with the Pomeron.

(b)  {\it \bf Strong Coupling and AdS-Graviton}: Prior to AdS/CFT,  property of the Pomeron has been explored  mostly from a perturbative approach. The advent of the AdS/CFT correspondence has provided a firmer foundation from which a non-perturbative treatment can now be carried out. 
 For instance, for elastic scattering,  in the {\it extreme limit of $\lambda\rightarrow \infty$},  the 2-to-2 amplitude can be represented by a Witten diagram of  {\bf a single graviton} propagating in $AdS_5$, schematically given in a factorizable form, 
$
A(s,t)  = \Phi_{13}*\widetilde {\cal K}_G * \Phi_{24} \; .
\label{eq:adsGravitonScheme}
$
where $ \Phi_{13}$ and $ \Phi_{24}$ represented two elastic vertex couplings to the graviton and $\widetilde {\cal K}_G$ is dominated by the ``$(++,--)$" component of the graviton propagator~\cite{Brower:2007qh}.  Since this corresponds to a spin-2 exchange, the dominant graviton kernel $\widetilde {\cal K}_G$ grows with a integral power, i.e., at fixed $t$, as  $s^2$.  

Similarly, double diffractive Higgs production will be dominated by a double-graviton exchange Witten diagram, leading to a similar factorizable expression for the production amplitude
 $
A(s,s_1, s_2, t_1, t_2)  = \Phi_{13}*  \widetilde{\cal K}_G *V_H*  \widetilde {\cal K}_G* \Phi_{24} \; ,
\label{eq:adsDoubleGravitonScheme}
$
where  $V_H$ is  a new Higgs production vertex. The central issue in a holographic description for diffractive Higgs production is the specification of this new vertex $V_H$.

(c) {\it From Graviton to Pomeron}:  It has been shown in \cite{Brower:2006ea}, for ${\cal N} = 4$ SUSY YM, 
the leading strong coupling Pomeron~\cite{Brower:2006ea,Brower:2007qh,Brower:2007xg} is at 
\be
j_0 = 2 - 2 /\sqrt{g^2N_c}    \; .
\label{eq:BPST-intercept}
\ee
which is ``lowered" from  $J =2$ as one decreases $\lambda=g^2N_c$.  In a realistic holographic approach to high energy scattering, one must work at $\lambda$ large but finite in order to account for  the Pomeron intercept  of the order $j_0\simeq 1.3$, leading to  (\ref{eq:adsPomeronScheme}) and (\ref{eq:adsDoublePomeronScheme}) for elastic and diffractive Higgs production respectively. Here the Pomeron kernel, $ \widetilde{\cal K}_P$,  has hard components due to near conformality in the UV and soft
Regge behavior in the IR~\footnote{Unlike the case of a graviton exchange in $AdS$, this Pomeron kernel contains both real and imaginary parts. For more discussion, see \cite{Brower:2006ea,Brower:2007xg,Brower:2012mk}.}.   

(d)  {\it Confinement}: Our  discussion above has been    formal since a CFT has no scale and  one needs to be more precise in defining the Regge limit.  With confinement deformation, both the dilaton and the transverse-traceless metric fluctuations become massive, leading to an infinite set of massive scalar and tensor glueballs respectively.  In particular, each glueball state can be described by a normalizable wave function $\Phi(z)$ in $AdS$. The weight factor   $\Phi_{ij}$ in the respective factorized representation for the elastic and Higgs amplitudes,  (\ref{eq:adsPomeronScheme}) and (\ref{eq:adsDoublePomeronScheme}), is given by $\Phi_{ij}(z) = e^{-2A(z)} \Phi_i(z)\Phi_j(z)$. In contrast, for amplitudes involving external currents, e.g., for DIS~\cite{Polchinski:2002jw,Brower:2010wf}, non-normalizable wave-functions will be used.

\section{Pomeron-Pomeron Fusion Vertex}\label{sec:Ingredient}
In a
perturbative approach, often dubbed as ``hard Pomeron", Higgs production can be viewed as gluon
fusion in the central rapidity region.  A Higgs can be produced at central rapidity by the double Regge Higgs vertex through
a heavy quark loop which in lowest order is a simple gluon fusion process, dominant for large parton x for the colliding gluons.
A more elaborate picture emerges as one tries to go  to the region of the softer (wee gluons) building up
double Regge regime~\footnote{In addition to  the Pomeron exchange contribution in these models must
subsequently be reduced by large Sudakov correction at the Higgs
vertex and by so called survival probability estimates for soft gluon
emission, again reflecting the view that double diffraction Higgs
production is intrinsically non-perturbative.}.
In the large $N_c$ there are no quark loop in the bulk of AdS space and since
the Higgs in the Standard Model only couples to quark via the
Yukawa interactions there appears to be a problem with strong coupling Higgs production
in leading $1/Nc$.  Fortunately the solution to this is to follow the standard procedure in Higgs phenomenology, which is to
integrate out the quark field replacing the Higgs coupling to the gauge operator $Tr[F^2]$. 

Consider the Higgs coupling to quarks via a Yukawa coupling,  and, for simplicity we will assume is dominated by the top quark. We will be more explicit in the next Section, and simply note here that, 
 after taking advantage of the scale separations between the QCD scale, i.e., the Higgs mass and the top quark mass,   $\Lambda_{qcd} \ll m_H \ll 2 m_t$, 
heavy quark decoupling allows one to replace the Yukawa coupling  by an effective interaction,
$
{\cal L} = \frac{\alpha_s g}{24 \pi M_W} F^a_{\mu\nu}F^{a \mu \nu} \phi_H
\label{eq:effectiveHiggsCoupling}
$,
by evaluating the two gluon Higgs triangle graph in leading order $O(M_H/m_t)$. 
Now the AdS/CFT dictionary simply requires that this be the source in the UV of the AdS dilaton field.  It follows, effectively, for Higgs production, we are required to work with a five-point amplitudes, one of the external leg involves a scalar dilaton current coupling to $Tr[F^2]$.  For diffractive Higgs production, in the supergravity limit, the Higgs vertex $V_H$ is given by a two-graviton-dilaton coupling, Fig. \ref{fig:cylindarHiggs}c.

We now must pause to realize that in any conformal theory the is no dimensional 
parameter to allow for such a  dimensionful two-graviton-dilaton coupling, $M^2 \phi h_{\mu\nu} h^{\mu \nu}$,  emerging  in an  expansion of the AdS gravity action  if scale invariance is maintained.  However since  QCD is not a conformal theory 
this is just one of many reasons to introduce {\it \bf conformal symmetry breaking}.  Many attempts have
been made to supplement this phenomenological Lagrangian with other fields such as
the  gauge fields for the light quark Goldstone modes to provide
a better holographic dual for QCD.  In principle enen at
leading order of large $N_c$ we should eventually
require  an infinite number of (higher spin) field in the bulk representation
to correspond the yet undiscovered 2-d sigma model for the world-sheet string theory
for  QCD.   Fortunately for the phenomenological
level at high energy, these details are non-essential.  To model an effective QCD background we will for the most
part introduce two modifications of the pure AdS background: (1) an IR hardwall cut-off 
beyond $z = 1/\Lambda_{qcd}$ to give confinement and  linear static quark potential at large distances and (2) 
a slow deformation in the UV ($z \rightarrow 0$) to model the logarithmic running 
for asymptotic freedom.  Both break conformal invariance, which
as we will argue is required to couple the two gravitons to the dilaton and produce a
Higgs in the central rapidity region.

After taking into account of finite $\lambda$ correction, the leading order Higgs production diagram at large $N_c$ can be schematically represented in Fig.~\ref{fig:cylindarHiggs}c, with each of the left- and right-cylinder representing a BPST Pomeron. It should be pointed out, just as in the case  of elastic scattering, it is necessary to consider higher order corrections, e.g., eikonal corrections. We will not do it here, but will address this issue in the conclusion section.  In what follows, we shall focus on the Pomeron-Pomeron fusion vertex in the strong coupling limit.

Finally it should be noted that
one critical missing ingredient of these ad hoc conformal breaking deformation of
the AdS geometry  in the UV and IR  is the
fact the spontaneous  breaking of pure Yang Mills ( and presumable  QCD
at large $N_c$), via ``dimensional transmutation'' 
eliminates the coupling, $\lambda$, as a free parameter. It is fixed 
via the beta function in terms of a single integration constant (sometime called  $\Lambda_{qcd}$ ) which provides the only mass scale.  Thus the logarithmic scale violation in the UV are tied to the same parameter giving confinement in the IR. All holographic modes of QCD to date introduce two mass scales and thus neglect this constraint. The solution to this problem also presumably awaits the
determination of the unique string theory for large $N_c$ QCD.

We are now in a position to focus on the issue of double diffractive
Higgs production from the perspective of String/Gauge duality, i.e.,
the Higgs vertex, $V_H$. It is important to stress that our general discussion  in moving from
single-Pomeron exchange processes, (\ref{eq:adsPomeronScheme}), to double-Pomeron exchange, (\ref{eq:adsDoublePomeronScheme}), applies equally well for both
diffractive glueball production and for Higgs production. The
difference lies in how to treat the new central vertex.  For the
production of a glueball, the vertex will be proportional to a
normalizable $AdS$ wave-function.  There will also be an overall factor
controlling the strength of coupling to the external states, e.g., the
Pomeron-Pomeron-glueball couplings.  For Higgs production, on the
other hand, the central vertex, $V_H$, involves a non-normalizable
bulk-to-boundary propagator, appropriate for a scalar external
current.This in turns leads to coupling to a Higgs scalar.  The
difference between these two cases parallels the situation for
four-point amplitudes in moving from proton-proton (p-p) elastic
scattering to electron-proton deep-inelastic scattering (e-p DIS). In
moving from p-p to DIS, one simply replaces one of the two pairs of
normalizable proton wave-functions  with a
pair of non-normalizable counterparts appropriate for conserved
external vector currents.

A Higgs scalar in the standard model couples exclusively to the quarks
via Yukawa coupling, which for simplicity we will assume is dominated
by the top quark, with   
$
{\cal L} = - \frac{g}{2 M_W} m_t \; \bar t(x) t(x) \phi_H(x).
\label{eq:HiggsCoupling2ttbar}
$
 Taking advantage of the scale separations between the QCD scale, the Higgs mass and the top quark mass,
$
\Lambda_{qcd} \ll m_H \ll 2 m_t
$,
heavy quark decoupling allows one to replace the Yukawa coupling by direct coupling of Higgs to gluons, which is treated  as an external source in the AdS dictionary.
Consequently $V_H$, in a coordinate representation,  is replaced by the vertex  for two AdS Pomerons fusing
at $(x'_{1\perp},z'_1)$ and $(x'_{2\perp},z'_2)$ and propagating this
disturbance to the $\bar t(x) t(x)$ scalar current at the boundary of
AdS.  The double diffractive Higgs vertex $V_H$ can then be obtained in a two-step process.

First, since the
Yukawa Higgs quark coupling is proportional to the quark mass, it is
dominated by the top quark. Assuming  $m_H \ll m_t$,  this can be
replaced by an effective interaction, (\ref{eq:effectiveHiggsCoupling}),
by evaluating the two gluon Higgs triangle graph in leading order $O(M_H/m_t)$.
Second, using the AdS/CFT dictionary, the external
  source for $F^a_{\mu \nu}F^a_{\mu \nu}(x)$ is placed at the
  AdS boundary ($z_0 \rightarrow 0$) connecting to the Pomeron fusion vertex
in the interior of $AdS_3$ at ${\bf b}_H=(x'_H,z'_H)$, by a scalar bulk-to-boundary propagator,
$K(x'_H - x_H,z'_H,z_0)$.

We are finally in the position to put all the pieces together. Although we eventually want to go to a coordinate representation in order to perform eikonal unitarization, certain simplification can be achieved more easily in working with the  momentum representation. The Higgs production amplitude, schematically given by (\ref{eq:adsDoublePomeronScheme}), can then be written explicitly as
 \bea
A(s,s_1,s_2, t_1,t_2)&\simeq &  \int dz_1 dz  dz_2\; \sqrt{-g_1}\sqrt{-g} \sqrt{-g_2}\;\Phi_{13} (z_1)     \nn
&\times&   \widetilde {\cal K}_P(s_1,t_1,z_1,z)  \; V_H(q^2 , z)\; \widetilde {\cal K}_P(s_2,t_2,z,z_2)\; \Phi_{24}(z_2)  \;.
\label{eq:adsDoublePomeronHiggs}
\eea
where  $q^2= -m_H^2$.   For this production vertex,  we will keep  it simple by  expressing it as 
\be
 V_H( q^2,z)= V_{PP\phi} K(q^2,z) L_H\; .
 \label{eq:HiggsVertex}
\ee
where  $K(q^2,z)$ is the conventionally normalized bulk to boundary propagator, $V_{PP\phi}$ serves as  an overall coupling from two-Pomeron to $F^2$, and $L$ is  the  conversion factor from $F^2$ to Higgs, i.e.,
$
L_H = L(-m_H^2)\simeq  \frac{\alpha_s g}{24 \pi M_W}.
$
We shall treat  the central vertex $  V_{PP\phi}$ as a constant, which follows from the super-gravity limit. This approximation gives an explicit
factorizable form for Higgs production.

The current version for the holographic Higgs amplitude~\cite{Brower:2012mk} involves 3  parameters: (1) the IR cut-off determined by the glueball mass, (2) the leading singularity in the $J$-plane determined~\footnote{In a true dual to QCD, there is no independent parameter for   the strong coupling, because of ``dimensional transmutation'', which fixes all dimensionful quantities   relative to the a single mass scale $\Lambda_{qcd}$, through the running   coupling constant. For instance, the glueball mass in units of $\Lambda_{qcd}$ is fixed and   computed in lattice computations. } by the 't Hooft parameter $\lambda$ and (3) the strength of the central vertex parameterized by the string coupling or Planck mass. A strategy must be provided in fixing  these parameters.

\section{Strategy for Phenomenological Estimates}
\label{sec:strategy}

As a first step in this direction, we ask how the central vertex, $V_H$, or equivalently,  $ V_{PP\phi}$,  via (\ref{eq:HiggsVertex}),  can be normalized, following  the approach of Kharzeev and Levin~\cite{Kharzeev:2000jwa} based on the analysis of trace anomaly. 
 We also show how
one can in principle use the elastic scattering to normalize the bare BPST Pomeron coupling to external protons and the 't Hooft 
coupling $g^2N_c$. As in the case of elastic scattering, it is pedagogically reasonable to begin by first treating the simplest case  of double-Pomeron exchange for Higgs production, i.e., without absorptive correction. We discuss how phenomenolgically reasonable simplifications can be made. This is followed by treating eikonal
corrections in the next section, which provides a means of  estimating the  all-important survival probability.

(a) {\it Continuation to Tensor Glueball Pole:}     Confinement deformation in AdS will lead to glueball states, e.g., the lowest tensor glueball state lying on the leading Pomeron trajectory~\cite{Brower:2000rp}.   There will also be scalar glueballs associated with the dilaton.  With scalar invariance broken, this  will also lead to non-vanishing couplings between a pair of tensor glueballs and scalar glueballs. 
In terms of the language of Witten diagram,  corresponds  to a non-vanishing graviton-graviton-dilaton coupling in the bulk, which  in turn leads to  $V_H\neq 0$.

Consider first the elastic amplitude. With confinement,
each Pomeron kernel  will contain a tensor glueball pole when $t$ goes on-shell. Indeed, the propagator for our Pomeron kernel can be expressed as a discrete sum over pole contributions.   That is, when $t\simeq  m_0^2$, where $m_0$ is the mass of the lightest tensor glueball, which lies on the leading Pomeron trajectory.  In this limit, the elastic amplitude then  takes on the expected  pole-dominated  form,
$
A(s,t) \simeq g_{13}\; \frac {s^2}{m^2_0-t} \; g_{24}
\label{eq:2to2KL}
$.
The external coupling  $g_{ij}$ is given by an overlap-integral over a product of three wave functions,   $\Phi_i(z)$, $ \Phi_j(z)$  and $  \phi_{G}(z)$, where $\phi_{G}(z)$ is the wave function for the tensor glueball. With the standard normalization, 
 $A(s,t)$ is dimensionless.

A similar analysis can also be carried out for   the Higgs production amplitude, Eq. (\ref{eq:adsDoublePomeronHiggs}). Note that the Pomeron kernel now appears twice, $\widetilde {\cal K}_P(s_1,t_1,z_1,z) $ and $\widetilde{\cal K}_P(s_2,t_2,z_2,z) $.   When nearing the respective tensor poles at $t_1\simeq  m_0^2$ and $ t_2\simeq  m_0^2$,  the amplitude can be expressed as
\bea
A(s,s_1,s_2, t_1,t_2)&\simeq&g_{13}\; \frac{  \Gamma_{GGH} \;  s^2  }{(t_1-m_0^2)(t_2-m_0^2)  }   \; g_{24}
   \label{eq:2to3KL}
\eea
\ignore{As for the elastic case, we have performed the $z_1$ and $z_2$ integrations,  and   have also made use of the fact that $s_1s_2\simeq \kappa\; s\simeq m_H^2 s $. }
Here $\Gamma_{GGH}$ is the effective on-shell glueball-glueball-Higgs coupling, which  can also be expressed as
$
 \Gamma_{GGH} =L_H F ( - m^2_H)
$,
where $L_H=    \frac{\alpha_s g}{24 \pi M_W}$ and     $ F $  is a  form factor
$
F(q^2) = \<G, ++,q_1| F^a_{\mu\nu} F^a_{\mu\nu} | G,--,q_2\>.
$ 
which can again be expressed as an overlapping integral involving the bulk-boundary propagator and the glue ball wave function.   What remains to be specified is the overall normalization, $F(0)$.

We next follow  D. Kharzeev and E. M. Levin  \cite{Kharzeev:2000jwa}, who noted  that, from the SYM side,  $F(q^2)$ at $q^2=0$, can be considered as the glueball condensate, leading directly to
\be
F(0) = \<G| F^a_{\mu \nu} F^{a \mu \nu} | G  \>  =-  \frac{4\pi  M_G^2}{3 \widetilde \beta }
\label{eq:FF}
\ee
where $\widetilde \beta = - b\alpha_s/(2 \pi)$, $b = 11 - 2n_f/3$, for $N_c=3$.  In what follows, we will use $n_f=3$.   Note that heavy quark contribution is not included in this  limit.  Since the conformal scale breaking is due the running coupling constant in QCD, 
there is apparently a mapping between QCD scale breaking and breaking of
the AdS background  in the IR, which  gives a finite mass to the glueball
and to give a non-zero contribution to the gauge condensate.

(b) {\it Extrapolation to  the Near-Forward limit:}   To apply the above result to the physical region, one needs to extrapolate from $t$ near the tensor pole to the physical region where $t_1,t_2\simeq 0$. 
 In addition, exponential  cutoff for $t_1$ and $t_2$ small also have to be taken into account.  For elastic scattering, it is customary to   express  the  amplitude in the near forward region as $ 
A(s,t) \simeq  e^{B_{eff}(s)\; t }\;  A(s,0)
$ 
where $B_{eff}(s)$ is a smoothly slowly increasing function of s, (we expect it to be logarithmic).
Similarly, we also assume, for $t_1<0$, $t_2<0$ and small, the Higgs production amplitude is also strongly damped so that
\be
A(s,s_1,s_2, t_1, t_2)  \simeq    e^{B'_{eff}(s_1) \; t_1/2}  e^{B'_{eff}(s_2)\; t_2 /2   }  \; A(s,s_1,s_2, t_1\simeq 0, t_2\simeq 0).
\label{eq:higgs}
\ee
   Although these steps are  conceptually straight forward, considerable details have to be spelled out in arriving a manageable expression, as done in \cite{Brower:2012mk}. One finds that 
\bea
\frac{d\sigma}{dy_H } &\simeq &(1/\pi) \times C  \times |   \Gamma_{GGH}(0)/\widetilde m^2|^2  \times \frac{\sigma( s)}{\sigma(m_H^2)} \times R^2_{el}(m_H \sqrt s )
\eea
Here,  $
R_{el}(s) = {\sigma_{el}}/{\sigma_{total}}  \simeq  \frac{ (1+\rho^2) \sigma_{total}(s) }{16\pi B_{eff} (s) }
$.   In this expression above, both $C$ and $\widetilde m^2$, like  $ m_0^2$,  are model dependent. It is nevertheless interesting to note that, since $\Gamma_{GGH}(0)\sim m_0^2$, the glueball mass scale also drops out, leaving a model-dependent ratio of order unity.  In deriving the result above, we have replaced $B'_{eff}$ by $B_{eff}$ where the difference is unimportant at high energy.  With $m_H\simeq 150$ GeV, $R_{el}$ can be taken to be  in the range $0.1$ to $0.2$. For $C\simeq 1$, we find  $\frac{d\sigma}{dy_H }\simeq .8\sim 1.2  \;\; {\rm pbarn}$.
This is of the same order as estimated in \cite{Kharzeev:2000jwa}.  However, as also pointed in \cite{Kharzeev:2000jwa}, this should be considered as an over-estimate. The major source of suppression will come from absorptive correction, which can lead to a central production cross section in the  fb   range. We turn to this next.

\section{Discussion}
\label{sec:discusion}

We conclude by  discussing how consideration of higher order contributions via an eikonal treatment leads to further suppression for the central Higgs production. Following by now established usage, the resulting production cross section can be expressed in terms of a ``survival probability". 

For elastic scattering,   the resulting  eikonal sum leads
to an impact representation for the 2-to-2 amplitude:
%
$
A(s,x^\perp - x'^\perp)= - 2i s \int
 dz \ dz'\ P_{13}(z) P_{24}(z')  \left [
e^{i\chi(s,x^\perp - x'^\perp, z,z')} - 1\right].
\label{eq:adseik1}
$
%
  The eikonal $\chi$, as  a function of $x_\perp- x'_\perp$, $z,
z'$ and $ s$, can be determined by matching the first order term in $\chi$
to the
single-Pomeron contribution.
This eikonal analysis can be extended directly to Higgs production. Since Higgs production is a small effect,  we find the leading order Higgs production amplitude, to all order in $\chi$, becomes
\bea
A_H(s_1, s_2,x^\perp -
x^\perp_H,x'^\perp -x^\perp_H, z_H) &=& 2 s \int
 dz \ dz'\ P_{13}(z) P_{24}(z')  \nn
 &\times&  \chi_H(s_1, s_2,x^\perp -
x^\perp_H,x'^\perp -x^\perp_H, z,z',z_H) \;    e^{i\chi(s,x^\perp - x'^\perp, z,z')}\nonumber
\eea
where $  \chi_H $ can be found  by matching in the limit $\chi\rightarrow 0$ with the Higgs production amplitude, (\ref{eq:adsDoublePomeronHiggs}),   in an impact representation. The net effect of eikonal sum is to introduce a phase factor 
$
e^{i\chi(s,x_\perp- x'_\perp,z,z')} $
 into the production amplitude. Due to its absorptive part, ${\rm Im}\; \chi >0$, this eikonal factor provides a strong suppression for central Higgs production.  
 
 The effect of this suppression is often expressed in terms of a ``Survival Probability", $\langle S \rangle$.   In a momentum representation, the  cross section for Higgs production per unit of
rapidity in the central region is
$
\frac{d \sigma_H(s,y_H)}{dy_H} = \frac{ 1\;  }{ \pi^3 (16\pi)^2  s^2 } \int d^2q_{1\perp}  d^2q_{2\perp} |A_H(s,y_H,
q_{1\perp},q_{2\perp})|^2  
$
where $y_H$ is the rapidity of the produced Higgs, $q_{1\perp} $ and $q_{2\perp} $ are transverse momenta of two outgoing fast leading particle in the frame where the momenta of  incoming particles are longitudinal.   ``Survival Probability" is conventionally defined by the ratio
$
\langle S \rangle\equiv \frac{  \int d^2q_{1\perp}  d^2q_{2\perp} |A_H(s,y_H,
q_{1\perp},q_{2\perp})|^2  }{ \int d^2q_{1\perp}  d^2q_{2\perp} |A^{(0)}_H(s,y_H,
q_{1\perp},q_{2\perp})|^2}
\label{eq:survival}
$,
where $A^{(0)}_H$ is the corresponding amplitude before eikonal suppression, e.g., given by Eq. (\ref{eq:adsDoublePomeronHiggs}).  
For simplicity, we shall also focus on the  mid-rapidity production, i.e., $y_H\simeq 0$ in the overall CM frame. In this case, $\langle S \rangle$ is a function of overall CM energy squared, $s$, or the equivalent total rapidity, $Y\simeq \log s$.   Evaluating the survival probability as given by (\ref{eq:survival}), though straight forward, is  often tedious.

To gain a qualitative estimate, let us consider   the  local limit where $z\simeq \bar z\simeq z_0$ and  $z'\simeq \bar z'\simeq z'_0$, with $z_0\simeq z_0'\simeq 1/\Lambda_{QCD}$. In this limit,  one finds that this suppression factor reduces to
$
e^{- 2\; {\rm Im}\; \chi (s,x_\perp,x'_\perp,z_0,z'_0)}
$,
where $ {\rm Im}\; \chi>0$ by unitarity. If follows that, in a super-gravity limit of strong coupling where the eikonal is strictly real, there will be no suppression and the survival probability is 1. Conversely,  the fact that phenomenologically a small survival probability is required provides  another evidence the need  to work in an intermediate region where $1< j_0<2$.  In this more realistic limit, ${\rm Im}\; \chi$ is large and cannot be neglected.   In particular,  it follows that the dominant region for diffractive Higgs production in pp scattering comes from  the region where
$
{\rm Im} \;\;\chi(s,x_\perp- x'_\perp,z,z') = O(1),
$
with $z\simeq z' =O(1/\Lambda_{qcd})$. Note that this  is precisely the edge of the ``disk region"  for p-p scattering.   In order to carry out a quantitative analysis, it is imperative that we learn the property of $\chi(s,\vec b, z)$ for $|\vec b|$ large. From our experience with pp scattering, DIS at HERA, etc., we know that confinement will play a crucial role. In pp scattering, since $z\simeq z' =O(1/\Lambda_{qcd})$, we expect this condition is reached at relatively low energy, as is the case for total cross section. It therefore plays a dominant role in determining the magnitude of diffractive Higgs production at LHC.  We will not discuss this issue here further; more pertinent discussions on how to determine $\chi(s,x_\perp- x'_\perp,z,z')$ when confinement is important can be found in Ref. \cite{Brower:2010wf}.

We have focussed here primarily  on central exclusive Higgs production from a holographic perspective. In such a treatment,  a non-perturbative Pomeron corresponds to exchanging a Reggeized Graviton in $AdS$. This approach can of course also be applied to central exclusive production of mesons~\cite{Anderson:2014jia}. Equally interesting is the possibility of further generalization of incorporation of Odderon exchanges~\cite{Brower:2009zz}. For instance, a  Pomeron-Odderon double-exchange can be used to discussion the production of ``axions", a subject which is under current investigation.



\end{document}